\documentstyle[12pt,bezier]{article}
\begin{document}
\newcommand{\de}{\delta}\newcommand{\ga}{\gamma}
\newcommand{\e}{\epsilon} \newcommand{\th}{\theta}\newcommand{\ot}{\otimes}
\newcommand{\ba}{\begin{array}} \newcommand{\ea}{\end{array}}
\newcommand{\beq}{\begin{equation}}\newcommand{\eeq}{\end{equation}}
\newcommand{\tmod}{{\cal T}}\newcommand{\amod}{{\cal A}}
\newcommand{\bemod}{{\cal B}}\newcommand{\cmod}{{\cal C}}
\newcommand{\dmod}{{\cal D}}\newcommand{\hmod}{{\cal H}}
\newcommand{\s}{\scriptstyle}\newcommand{\tr}{{\rm tr}}
\newcommand{\einsop}{{\bf 1}}
\title{Bethe ansatz solution of a closed spin 1 XXZ Heisenberg chain
with quantum algebra symmetry}
\author{Jon Links\dag\thanks{e-mail jrl@maths.uq.oz.au}, \quad
Angela Foerster\ddag\thanks{ e-mail angela@if.ufrgs.br} \quad and
\quad Michael Karowski\ddag\thanks{e-mail karowski@physik.fu-berlin.de}}
\date{ \dag Department of Mathematics, University of
Queensland, \\ Queensland, 4072,  Australia \\
\vspace{0.35cm}
\ddag Institut f\"ur Theoretische Physik, Freie Universt\"at
Berlin, \\ Arnimmallee 14, Berlin, Germany}
\maketitle
\begin{abstract}
A quantum algebra invariant integrable closed spin 1 chain
is introduced and analysed in detail. The Bethe ansatz
equations as well as the energy eigenvalues of the model
are obtained. The highest weight property of the Bethe
vectors with respect to $U_q(sl(2)) $ is proved.
\end{abstract}


Running Title - Closed spin 1 $XXZ$ chain 

\newpage
\section*{I. Introduction }

The Quantum Inverse Scattering Method has proved to be a powerful
procedure in the analysis of one-dimensional integrable quantum chains
or two-dimensional lattice models of statistical mechanics (e.g. see
\cite{KIB}).
Central to this formalism is the Yang-Baxter equation whose solutions
are sufficient to guarrantee integrability of the associated model. The
advent of quantum algebras \cite{J,D} provided a systematic treatment to obtaining
solutions of the Yang-Baxter equation. However the most common approach
to the QISM, which was to impose periodic boundary conditions, was
quickly realized to be incompatible with the quantum algebra symmetry of
the model. Several authors were able to overcome this problem by working
with a model on an open chain \cite{Des,Gr2,For2,Gr1,Rui}.
This practice made the Bethe
ansatz solutions of such models more difficult and in some instances
only postulated solutions are available \cite{Nep2}.

More recently, it has been demonstated that it is in fact possible to
construct closed chain models with preservation of quantum algebra
symmetry \cite{Kar,GPPR,F,Lima}. Significantly the $U_q(sl(2))$ invariant
closed spin 1/2 $XXZ$ model was
shown to be connected with a lattice quantization of the Liouville
model \cite{FT}. The algebraic Bethe ansatz solutions of such models
showed that the closed chain quantum algebra invariant case was not
fraught with the same difficulties which were faced in the instances of
open chains. The existence of such symmetry makes available results such
as the highest weight property of the Bethe states for the fundamental
representation of $U_q(sl(n))$, and furthermore a characterization of
``good'' and ``bad'' states in terms of $q$-dimensions when $q$ takes
values of roots of unity \cite{Kar1}. Initially, just  quantum algebra
invariant closed chains of the Hecke algebra type were analysed.
It was subsequently shown \cite{Jonangi} that a more general
formulation existed.

Here we wish to expand on the knowledge of closed chain quantum algebra
invariant models by undertaking a detailed study of the $U_q(sl(2))$
invariant spin 1 model. Integrable spin 1 models built from a $U_q(sl(2))$
invariant $R$-matrix have already been the subject of
some analysis  \cite{BMNR,Mnr}. Our study of the closed chain $U_q(sl(2))$ 
invariant spin 1 model exposes new
mathematical aspects not present in the previously studied models
\cite{Kar,GPPR,F}; viz. the model is not of Hecke algebra type and it is an
example of a higher spin system where the most natural approach to the
Bethe ansatz solution is to use a transfer matrix defined on 
an auxiliary space different from the local quantum space. We then find the
eigenvalues of the transfer matrix whose auxiliary space is isomorphic
to the local quantum space following the method of Babujian and
Tsvelick \cite{bab2}. The need to use two transfer matrices defined on
different auxiliary spaces means working with more than one solution of
the Yang-Baxter equation. Throughout we will follow the notation of
\cite{bab2,bab1} in distinguishing the spaces on which the various 
operators act.
Specifically we use the symbol $\sigma$ to denote action on the spin 1/2
space and $s$ for action on the spin 1 space. 

The paper is organized as follows. In section 2 we define some
basic quantities, e.g., $R$ matrices, monodromy and transfer
matrices. A quantum algebra invariant closed spin-1 chain is
introduced and its relation with one of the transfer matrices
is presented. In section 3 the system is analysed through a
combination of the techniques developed to handle with
quantum algebra invariant closed chains \cite{Kar}
and higher-spin chains \cite{Mnr} and the Bethe
ansatz equations as well as the energy eigenvalues
of the model are obtained.
In section 4 we show that the Bethe vectors are highest
weight vectors with respect to $U_q(sl(2)) $.
We also argue that use of the $U_q(sl(2))$ generators allows use to
generate a complete set of states for the model.
A summary of our main results is presented in section 5.

\section*{II. The model}
We begin by recalling the $R$-matrix for the spin-1/2 chain
\begin{equation}
\label{1}
{\footnotesize
\phantom{0}_{\sigma \sigma } R^{\beta_1 \beta_2}_{\alpha_1 \alpha_2}(x)=
\ba{c}
\unitlength=0.50mm
 \begin{picture}(20.,25.)
\put(-3.,2.){\makebox(0.,0.){$\s x$}}
\put(0.,-4.){\makebox(0.,0.){$\s \alpha_2 $}}
\put(23.,2.){\makebox(0.,0.){$\s 1$}}
\put(20.,-4.){\makebox(0.,0.){$\s \alpha_1 $}}
\put(0.,24.){\makebox(0.,0.){$\s \beta_1 $}}
\put(20.,24.){\makebox(0.,0.){$\s \beta_2 $}}
\put(0.,20){\vector(-1,1){1.}}
\put(20.,20){\vector(1,1){1.}}
\put(0.00,0.00){\circle{2.0}}
\put(1.00,1.00){\circle{2.0}}
\put(2.00,2.00){\circle{2.0}}
\put(3.00,3.00){\circle{2.0}}
\put(4.00,4.00){\circle{2.0}}
\put(5.00,5.00){\circle{2.0}}
\put(6.00,6.00){\circle{2.0}}
\put(7.00,7.00){\circle{2.0}}
\put(8.00,8.00){\circle{2.0}}
\put(9.00,9.00){\circle{2.0}}
\put(10.00,10.00){\circle{2.0}}
\put(11.00,11.00){\circle{2.0}}
\put(12.00,12.00){\circle{2.0}}
\put(13.00,13.00){\circle{2.0}}
\put(14.00,14.00){\circle{2.0}}
\put(15.00,15.00){\circle{2.0}}
\put(16.00,16.00){\circle{2.0}}
\put(17.00,17.00){\circle{2.0}}
\put(18.00,18.00){\circle{2.0}}
\put(19.00,19.00){\circle{2.0}}
\put(20.00,0.00){\circle{2.0}}
\put(19.00,1.00){\circle{2.0}}
\put(18.00,2.00){\circle{2.0}}
\put(17.00,3.00){\circle{2.0}}
\put(16.00,4.00){\circle{2.0}}
\put(15.00,5.00){\circle{2.0}}
\put(14.00,6.00){\circle{2.0}}
\put(13.00,7.00){\circle{2.0}}
\put(12.00,8.00){\circle{2.0}}
\put(11.00,9.00){\circle{2.0}}
\put(10.00,10.00){\circle{2.0}}
\put(9.00,11.00){\circle{2.0}}
\put(8.00,12.00){\circle{2.0}}
\put(7.00,13.00){\circle{2.0}}
\put(6.00,14.00){\circle{2.0}}
\put(5.00,15.00){\circle{2.0}}
\put(4.00,16.00){\circle{2.0}}
\put(3.00,17.00){\circle{2.0}}
\put(2.00,18.00){\circle{2.0}}
\put(1.00,19.00){\circle{2.0}}
\end{picture}
\ea =
\frac1{\phantom{0}_{\sigma \sigma }a}
\pmatrix{\phantom{0}_{\sigma \sigma }a&0&|& 0&0\cr
         0&\phantom{0}_{\sigma \sigma }b&|& \phantom{0}_{\sigma \sigma }c_-&0\cr
        -&-& &  -&-&-\cr
         0&\phantom{0}_{\sigma \sigma }c_+&|& \phantom{0}_{\sigma \sigma }b&0\cr
         0&0&|&  0&\phantom{0}_{\sigma \sigma }a\cr} } \, \, \, ,
\end{equation}
with
\begin{equation}
\label{2}
\phantom{0}_{\sigma \sigma }a = x q - { 1 \over x q } ,\quad
\phantom{0}_{\sigma \sigma }b = x - { 1 \over x } ,\quad
\phantom{0}_{\sigma \sigma }c_+ = x \left( q - { 1 \over q} \right), \quad
\phantom{0}_{\sigma \sigma }c_- = { 1 \over x } \left(q - { 1 \over q} \right),\quad
\end{equation}
which acts in the tensor product of two 2-dimensional auxiliary
spaces ${\bf C}^2 \otimes {\bf C}^2$. Above
$\alpha_1$ , $\alpha_2$ ( $\beta_1$ and $\beta_2$ ) are column ( row
) indices
running from 1 to 2.

For the spin-1 chain the $R$-matrix is given by \cite{Zam,KR}
\begin{eqnarray}
\label{3}
\lefteqn{
\phantom{0}_{s s }R^{j_1 j_2}_{i_1 i_2}(x)=
\ba{c}
\unitlength=0.50mm
\begin{picture}(20.,25.)
\put(-3.,2.){\makebox(0.,0.){$\s x$}}
\put(0.,-4.){\makebox(0.,0.){$\s i_2 $}}
\put(23.,2.){\makebox(0.,0.){$\s 1$}}
\put(20.,-4.){\makebox(0.,0.){$\s i_1 $}}
\put(0.,24.){\makebox(0.,0.){$\s j_1 $}}
\put(20.,24.){\makebox(0.,0.){$\s j_2 $}}
\put(20.,0){\vector(-1,1){20.}}
\put(0.,0){\vector(1,1){20.}}
\put(0.,0.){\line(1,1){20.}}
\put(0.,20.){\line(1,-1){20.}}
\end{picture}
\ea}\\[5mm]\nonumber
&=&\frac1{_{s s }g}
\pmatrix{_{s s }g&0&0&|&  0&0&0&|&  0&0&0\cr
 0&_{s s }a&0&|&  _{s s }c_-&0&0&|&  0&0&0\cr
 0&0&_{s s }b&|&  0&_{s s }d_-&0&|&  _{s s }e_-&0&0\cr
 -&-&-& &  -&-&-& &  -&-&-\cr
 0&_{s s }c_+&0&|&  _{s s }a&0&0&|&  0&0&0\cr
 0&0&_{s s }d_+&|&  0&_{s s }f&0&|&  _{s s }d_-&0&0\cr
 0&0&0&|&  0&0&_{s s }a&|&  0&_{s s }c_-&0\cr
 -&-&-& &  -&-&-& &  -&-&-\cr
 0&0&_{s s }e_+&|&  0&_{s s }d_+&0&|&  _{s s }b&0&0\cr
 0&0&0&|&  0&0&_{s s }c_+&|&  0&_{s s }a&0\cr
 0&0&0&|&  0&0&0&|&  0&0&_{s s }g\cr},
\end{eqnarray}
where
\begin{eqnarray}
\label{4}
\phantom{0}_{s s}g &=& x q^2 - { 1 \over x q^2 } ,\quad
\phantom{0}_{s s}a = x - { 1 \over x } ,\quad
\phantom{0}_{s s}b = \left( x - { 1 \over x } \right) \left( \frac{x^2 - q^2}{x^2 q^2 -1} \right),
\nonumber \\
\phantom{0}_{s s }c_- &=& { 1 \over x } \phantom{0}_{s s }c, \quad
\phantom{0}_{s s}c_+ = x \phantom{0}_{s s }c, ,\quad
\phantom{0}_{s s }c, = \left( q^2 - { 1 \over  {q^2} }\right), \quad
\phantom{0}_{s s}f =  \phantom{0}_{s s}a \, + \,  \phantom{0}_{s s}e,
\nonumber \\
\phantom{0}_{s s}d_- &=& { 1 \over x } \phantom{0}_{s s}d ,\quad
\phantom{0}_{s s}d_+ = x  \phantom{0}_{s s}d , \quad
\phantom{0}_{s s}d = \left( \frac{x q}{x^2 q^2 -1} \right)
\left( x - { 1 \over x } \right)
 \left( q^2 - { 1 \over  {q^2} } \right), \\
\phantom{0}_{s s}e_- &=& { 1 \over x^2 }\phantom{0}_{s s}e ,\quad
\phantom{0}_{s s}e+ = x^2 \phantom{0}_{s s} e, \quad
\phantom{0}_{s s}e = \left( \frac{x q}{x^2 q^2 -1} \right)\left( q - { 1 \over q } \right) \left( q^2 - { 1 \over  {q^2} }\right),
\nonumber
\end{eqnarray}
and it acts in ${\bf C}^3 \otimes {\bf C}^3$, with ${\bf C}^3$ a
3-dimensional auxiliary space.

For later convenience we also introduce an $R$-matrix which
acts on ${\bf C}^2 \otimes {\bf C}^3$ \cite{KR}
\begin{equation}
\label{5}
{\footnotesize
\phantom{0}_{\sigma s }R^{\beta j}_{\alpha i}(x)=
\ba{c}
\unitlength=0.50mm
\begin{picture}(20.,25.)
\put(-3.,2.){\makebox(0.,0.){$\s x$}}
\put(0.,-4.){\makebox(0.,0.){$\s i$}}
\put(23.,2.){\makebox(0.,0.){$\s 1$}}
\put(20.,-4.){\makebox(0.,0.){$\s \alpha $}}
\put(0.,24.){\makebox(0.,0.){$\s \beta $}}
\put(20.,24.){\makebox(0.,0.){$\s j $}}
\put(0.,20){\vector(-1,1){1.}}
\put(20.,20){\vector(1,1){1.}}
\put(20.00,0.00){\circle{2.0}}
\put(19.00,1.00){\circle{2.0}}
\put(18.00,2.00){\circle{2.0}}
\put(17.00,3.00){\circle{2.0}}
\put(16.00,4.00){\circle{2.0}}
\put(15.00,5.00){\circle{2.0}}
\put(14.00,6.00){\circle{2.0}}
\put(13.00,7.00){\circle{2.0}}
\put(12.00,8.00){\circle{2.0}}
\put(11.00,9.00){\circle{2.0}}
\put(10.00,10.00){\circle{2.0}}
\put(9.00,11.00){\circle{2.0}}
\put(8.00,12.00){\circle{2.0}}
\put(7.00,13.00){\circle{2.0}}
\put(6.00,14.00){\circle{2.0}}
\put(5.00,15.00){\circle{2.0}}
\put(4.00,16.00){\circle{2.0}}
\put(3.00,17.00){\circle{2.0}}
\put(2.00,18.00){\circle{2.0}}
\put(1.00,19.00){\circle{2.0}}
\put(0.,0.){\line(1,1){20.}}
\end{picture}
\ea =\frac1{\phantom{0}_{\sigma s }a}
\pmatrix{\phantom{0}_{\sigma s }a&0&0&|&  0&0&0\cr
               0&\phantom{0}_{\sigma s }b&0&|&  \phantom{0}_{\sigma s }d_-&0&0\cr
               0&0&\phantom{0}_{\sigma s }c&|&  0&\phantom{0}_{\sigma s }d_-&0\cr
               -&-&-& &  -&-&-\cr
               0&\phantom{0}_{\sigma s }d_+&0&|&  \phantom{0}_{\sigma s }c&0&0\cr
               0&0&\phantom{0}_{\sigma s }d_+&|&  0&\phantom{0}_{\sigma s }b&0\cr
               0&0&0&|& 0&0&\phantom{0}_{\sigma s }a\cr}} \, \, \, ,
\end{equation}
where
\begin{eqnarray}
\label{6}
\phantom{0}_{\sigma s }a &=& x q^{3/2} - { 1 \over x q^{3/2} } ,\quad
\phantom{0}_{\sigma s}b =  x q^{1/2} - { 1 \over x q^{1/2} } ,\quad
\phantom{0}_{\sigma s}c =  {x\over q^{1/2}} - { q^{1/2} \over x } ,\quad \nonumber \\
\phantom{0}_{\sigma s }d_- &=& {1 \over  x} \phantom{0}_{\sigma s } d  ,\quad
\phantom{0}_{\sigma s}d_+ = x \phantom{0}_{\sigma s }d ,\quad
\phantom{0}_{\sigma s}d =  { \sqrt{ \left( q - { 1 \over q } \right) \left( q^2 - { 1 \over  {q^2} }\right)}} ,
\end{eqnarray}

These $R$-matrices satisfy the following properties
\begin{itemize}
\item Yang-Baxter equations
\begin{equation}
\label{8}
R^{\alpha^{\prime\prime} \beta^{\prime\prime} }_{\alpha^\prime
\beta^\prime}(x/y)
R^{\alpha^\prime \gamma^{\prime\prime} }_{ \alpha \phantom{0}
\gamma^\prime }(x)
R^{\beta^\prime \gamma^\prime}_{\beta \phantom{0} \gamma}(y) =
R^{\beta^{\prime\prime} \gamma^{\prime\prime} }_{\beta^\prime
\gamma^\prime}(y)
R^{\alpha^{\prime\prime} \gamma^\prime}_{\alpha^\prime \phantom{0}
\gamma}(x)
R^{\alpha^\prime \beta^\prime}_{\alpha \phantom{0} \beta}(x/y) \,
\, .
\end{equation}
\item
generalized Cherednik reflection property \cite{Cher}
\begin{equation}
\label{9}
R^{\alpha \phantom{0}  \beta \phantom{0}
}_{\alpha^\prime\beta^\prime  }(x)
(R^{-1})^{\alpha^\prime\beta^\prime}_{\gamma \phantom{0}  \delta
\phantom{0}} (y^{-1})=
R^{\alpha\phantom{0}  \beta \phantom{0}  }_{\alpha^\prime
\beta^\prime }(y) (R^{-1})^{\alpha^\prime \beta^\prime  
}_{\gamma \phantom{0} \delta \phantom{0}  }(x^{-1})
\end{equation}
\item crossing unitarity \cite{lg}   
\begin{equation}
(R^{t_1})^{\alpha\beta}_{\alpha'\beta'}(x\eta)K^{\alpha'}_{\alpha''}
((R^{-1})^{t_1})^{\alpha''\beta'}_{\gamma'\delta}(x)(K^{-1})^{\gamma'}_{
\gamma}=\delta^{\alpha}_{\gamma}\delta^{\beta}_{\delta}.
\label{cross}\end{equation} 
\end{itemize}
where $t_1$ denotes matrix
transposition in the first space, $\eta$ is a crossing parameter and
$K=K^t$ is the crossing matrix. Explicit forms for $K$ are given below.
We remark that eq. (\ref{9}) is the natural generalization of
Cherednik's reflection property to the case where the $R$-matrix acts on
two non-isomorphic spaces. 
 
Let us now introduce the ``doubled''  monodromy matrix
$ {\phantom{0}}_{s s} \cal{U}$
\begin{eqnarray}
\label{10}
\lefteqn{{\phantom{0}}_{s s}{\cal{U}}^{l \{ j \} }_{k \{ i \} }(x)\ =
\ba{c}
\unitlength=0.50mm
\begin{picture}(95.,49.)
\put(46.,15.){\makebox(0,0)[cc]{$\cdots$}}
\put(21.,-3.){\makebox(0,0)[cc]{$\s i_1$}}
\put(31.,-3.){\makebox(0,0)[cc]{$\s i_2$}}
\put(71.,-3.){\makebox(0,0)[cc]{$\s i_L$}}
\put(21.,47.){\makebox(0,0)[cc]{$\s j_1$}}
\put(31.,47.){\makebox(0,0)[cc]{$\s j_2$}}
\put(71.,47.){\makebox(0,0)[cc]{$\s j_L$}}
\put(70.,0.){\vector(0,1){45.}}
\put(30.,0.){\vector(0,1){45.}}
\put(20.,0.){\vector(0,1){45.}}
\put(73.,35.){\vector(1,0){8.}}
\put(57.,15.){\line(1,0){25.}}
\put(47.,35.){\makebox(0,0)[cc]{$\cdots$}}
\put(15.,15.){\line(1,0){21.}}
\put(30.,15.){\vector(-1,0){15.}}
\put(22.,35.){\line(1,0){6.}}
\put(32.,35.){\line(1,0){6.}}
\put(57.,35.){\line(1,0){8.}}
\put(80.,35.){\line(0,0){0.}}
\put(15.,35.){\line(1,0){4.}}
\put(5.,25.){\line(1,1){10.}}
\put(5.,25.){\line(1,-1){10.}}
\put(5.,25.00){\circle*{2.}}
\put(86.,15.){\makebox(0,0)[cc]{$\s k$}}
\put(86.,35.){\makebox(0,0)[cc]{$\s l$}}
\end{picture}
\ea}\\
&=&
{\phantom{0}}_{s s}{R_-}^{ l_2\,j_1}_{ l^\prime\, j_1^\prime}
{\phantom{0}}_{s s}{R_-}^{l_3\, j_2 }_{l_2\, j_2^\prime }
\dots
{\phantom{0}}_{s s}{R_-}^{l\, j_L }_{l_L\, j_L^\prime }
{\phantom{0}}_{s s}R^{l^\prime \, j_1^\prime}_{k_2 \, i_1}(1/x)
\nonumber 
{\phantom{0}}_{s s}R^{k_2 \, j_2^\prime}_{k_3 \, i_2}(1/x)
\dots
{\phantom{0}}_{s s}R^{k_L \, j_L^\prime}_{k \, i_L}(1/x)\ ,
\end{eqnarray}
which acts in the tensor product of a three-dimensional auxiliary
space and a quantum space ${\bf C}^3 \otimes {\bf C}^{3L}$ and
can be regarded as a $3\times3$ matrix of matrices acting in
the quantum space
\begin{equation}
\label{11}
\phantom{0}_{s s}{\cal{U}}^{l}_{k}(x) =
\pmatrix{\phantom{0}_{s s}{\cal{U}}^{1}_{1} \,&
         \phantom{0}_{s s}{\cal{U}}^{1}_{2} \,&
         \phantom{0}_{s s}{\cal{U}}^{1}_{3} \, \cr
         \phantom{0}_{s s}{\cal{U}}^{2}_{1} \, &
         \phantom{0}_{s s}{\cal{U}}^{2}_{2} \, &
         \phantom{0}_{s s}{\cal{U}}^{2}_{3} \, \cr
         \phantom{0}_{s s}{\cal{U}}^{3}_{1} \, &
         \phantom{0}_{s s}{\cal{U}}^{3}_{2} \, &
         \phantom{0}_{s s}{\cal{U}}^{3}_{3} \,
          }  \, \, \, ,
\end{equation}
Above the constant $\phantom{0}_{s s} R$-matrix is defined as
\begin{equation}
\label{12}
{\phantom{0}_{s s }R_- }
= -\lim_{x \rightarrow 0}
\phantom{0}_{s s }R^{-1}(x) \, = \,
\ba{c}
\unitlength=0.3mm
\begin{picture}(30.,30.)
\put(30.,0.){\vector(-1,1){30.}}
\put(18.,18.){\vector(1,1){12.}}
\put(0.,0.){\line(1,1){12.}}
\end{picture}
\ea\ ,
\end{equation}

For later convenience we also define the auxiliary ``doubled''
monodromy matrix
\begin{eqnarray}
\label{13}
\lefteqn{{\phantom{0}}_{\sigma s}{\cal{U}}^{\beta \{j\}}_{\alpha\{i\}}(x)\ =
\ba{c}
\unitlength=0.50mm
\begin{picture}(95.,49.)
\put(45.,15.){\makebox(0,0)[cc]{$\cdots$}}
\put(21.,-3.){\makebox(0,0)[cc]{$\s i_1$}}
\put(31.,-3.){\makebox(0,0)[cc]{$\s i_2$}}
\put(71.,-3.){\makebox(0,0)[cc]{$\s i_L$}}
\put(21.,47.){\makebox(0,0)[cc]{$\s j_1$}}
\put(31.,47.){\makebox(0,0)[cc]{$\s j_2$}}
\put(71.,47.){\makebox(0,0)[cc]{$\s j_L$}}
\put(86.,15.){\makebox(0,0)[cc]{$\s \alpha$}}
\put(70.,0.){\vector(0,1){45.}}
\put(30.,0.){\vector(0,1){45.}}
\put(20.,0.){\vector(0,1){45.}}
\put(80.,35.){\vector(1,0){1.}}
\put(57.00,35.00){\circle{2.}}
\put(58.00,35.00){\circle{2.}}
\put(59.00,35.00){\circle{2.}}
\put(60.00,35.00){\circle{2.}}
\put(61.00,35.00){\circle{2.}}
\put(62.00,35.00){\circle{2.}}
\put(63.00,35.00){\circle{2.}}
\put(64.00,35.00){\circle{2.}}
\put(65.00,35.00){\circle{2.}}
\put(66.00,35.00){\circle{2.}}
\put(67.00,35.00){\circle{2.}}
\put(68.00,35.00){\circle{2.}}
\put(72.00,35.00){\circle{2.}}
\put(73.00,35.00){\circle{2.}}
\put(74.00,35.00){\circle{2.}}
\put(75.00,35.00){\circle{2.}}
\put(76.00,35.00){\circle{2.}}
\put(77.00,35.00){\circle{2.}}
\put(78.00,35.00){\circle{2.}}
\put(57.00,15.00){\circle{2.}}
\put(58.00,15.00){\circle{2.}}
\put(59.00,15.00){\circle{2.}}
\put(60.00,15.00){\circle{2.}}
\put(61.00,15.00){\circle{2.}}
\put(62.00,15.00){\circle{2.}}
\put(63.00,15.00){\circle{2.}}
\put(64.00,15.00){\circle{2.}}
\put(65.00,15.00){\circle{2.}}
\put(66.00,15.00){\circle{2.}}
\put(67.00,15.00){\circle{2.}}
\put(68.00,15.00){\circle{2.}}
\put(69.00,15.00){\circle{2.}}
\put(70.00,15.00){\circle{2.}}
\put(71.00,15.00){\circle{2.}}
\put(72.00,15.00){\circle{2.}}
\put(73.00,15.00){\circle{2.}}
\put(74.00,15.00){\circle{2.}}
\put(75.00,15.00){\circle{2.}}
\put(76.00,15.00){\circle{2.}}
\put(77.00,15.00){\circle{2.}}
\put(78.00,15.00){\circle{2.}}
\put(79.00,15.00){\circle{2.}}
\put(80.00,15.00){\circle{2.}}
\put(45.,35.){\makebox(0,0)[cc]{$\cdots$}}
\put(86.,35.){\makebox(0,0)[cc]{$\s \beta$}}
\put(15.00,15.00){\circle{2.}}
\put(16.00,15.00){\circle{2.}}
\put(17.00,15.00){\circle{2.}}
\put(18.00,15.00){\circle{2.}}
\put(19.00,15.00){\circle{2.}}
\put(20.00,15.00){\circle{2.}}
\put(21.00,15.00){\circle{2.}}
\put(22.00,15.00){\circle{2.}}
\put(23.00,15.00){\circle{2.}}
\put(24.00,15.00){\circle{2.}}
\put(25.00,15.00){\circle{2.}}
\put(26.00,15.00){\circle{2.}}
\put(27.00,15.00){\circle{2.}}
\put(28.00,15.00){\circle{2.}}
\put(29.00,15.00){\circle{2.}}
\put(30.00,15.00){\circle{2.}}
\put(31.00,15.00){\circle{2.}}
\put(32.00,15.00){\circle{2.}}
\put(33.00,15.00){\circle{2.}}
\put(34.00,15.00){\circle{2.}}
\put(35.00,15.00){\circle{2.}}
\put(36.00,15.00){\circle{2.}}
\put(37.00,15.00){\circle{2.}}
\put(15.00,35.00){\circle{2.}}
\put(16.00,35.00){\circle{2.}}
\put(17.00,35.00){\circle{2.}}
\put(18.00,35.00){\circle{2.}}
\put(22.00,35.00){\circle{2.}}
\put(23.00,35.00){\circle{2.}}
\put(24.00,35.00){\circle{2.}}
\put(25.00,35.00){\circle{2.}}
\put(26.00,35.00){\circle{2.}}
\put(28.00,35.){\vector(1,0){1.}}
\put(32.00,35.00){\circle{2.}}
\put(33.00,35.00){\circle{2.}}
\put(34.00,35.00){\circle{2.}}
\put(35.00,35.00){\circle{2.}}
\put(36.00,35.00){\circle{2.}}
\put(37.00,35.00){\circle{2.}}
\put(39.00,35.){\vector(1,0){1.}}
\put(5.00,25.){\circle*{2.5}}
\put(5.00,25.00){\circle{2.}}
\put(6.00,26.00){\circle{2.}}
\put(7.00,27.00){\circle{2.}}
\put(8.00,28.00){\circle{2.}}
\put(9.00,29.00){\circle{2.}}
\put(10.00,30.00){\circle{2.}}
\put(11.00,31.00){\circle{2.}}
\put(12.00,32.00){\circle{2.}}
\put(13.00,33.00){\circle{2.}}
\put(14.00,34.00){\circle{2.}}
\put(15.00,35.00){\circle{2.}}
\put(5.00,25.00){\circle{2.}}
\put(6.00,24.00){\circle{2.}}
\put(7.00,23.00){\circle{2.}}
\put(8.00,22.00){\circle{2.}}
\put(9.00,21.00){\circle{2.}}
\put(10.00,20.00){\circle{2.}}
\put(11.00,19.00){\circle{2.}}
\put(12.00,18.00){\circle{2.}}
\put(13.00,17.00){\circle{2.}}
\put(14.00,16.00){\circle{2.}}
\put(15.00,15.00){\circle{2.}}
\end{picture}
\ea}\\\nonumber
&=&
_{\sigma s}{R_-}^{\beta_2 \, j_1 }_{
\alpha^\prime \, j_1^\prime} \,
\,_{\sigma s}{R_-}^{\beta_3 \, j_2 }_{
\beta_2 \, j_2^\prime} \,
\dots
\,_{\sigma s}{R_-}^{\beta \, j_L }_{
\beta_L \, j_L^\prime} \,
\,_{\sigma s}R^{\alpha^\prime \, j_1^\prime}_{
\alpha_2 \, i_1}(1/x) \,
\,_{\sigma s}R^{\alpha_2 \, j_2^\prime}_{\alpha_3 \, i_2}(1/x)
\dots
\,_{\sigma s}R^{\alpha_L \, j_L^\prime}_{\alpha \, \, i_L}(1/x)\ ,
\end{eqnarray}
where  ${\phantom{0}}_{\sigma s}{R_-}$ corresponds to the leading term
in the limit of the matrix $ {\phantom{0}}_{\sigma s}R^{-1}(x)$ for $x
\rightarrow 0 $, analogously to
$ {\phantom{0}}_{s s} R_- $ (see eq. (\ref{12}) )
It acts on
${\bf C}^2 \otimes {\bf C}^{3L}$ and can be represented
as a $2\times2$ matrix in the auxiliary space whose entries are
matrices acting in the quantum space
\begin{equation}
\label{14}
{\phantom{0}_{\sigma s}\cal{U}}^{\beta}_{\alpha }(x) =
\pmatrix{A\, &
         B\, \cr
         C\, &
         D\, \cr}  \, \, \, ,
\end{equation}
Using equations (\ref{8},\ref{9}) we can prove that the following Yang-Baxter
equations are fulfilled
\begin{equation}
\label{15}
\footnotesize{
{\phantom{0}}_{\sigma \sigma}R^{\alpha \phantom{0} \beta
\phantom{0}}_{\alpha^\prime \beta^\prime}(y/x)
{\phantom{0}}_{\sigma s}{\cal U}^{\beta^\prime}_{\gamma^\prime}(x)
{\phantom{0}}_{\sigma \sigma}{R_-}^{\alpha^\prime
\gamma^\prime}_{\delta^\prime \phantom{0}\alpha \phantom{0}}
{\phantom{0}}_{\sigma s}{\cal U}^{\delta^\prime}_{\delta}(y) =
{\phantom{0}}_{\sigma s}{\cal U}^{\alpha}_{\alpha^\prime}(y)
{\phantom{0}}_{\sigma \sigma}{R^+}^{\alpha^\prime \phantom{0}
\beta}_{\delta^\prime\beta^\prime}
{\phantom{0}}_{\sigma s}{\cal U}^{\beta^\prime}_{\gamma^\prime}(x)
{\phantom{0}}_{\sigma \sigma} (R^{-1})^{\delta^\prime \gamma^\prime}_{\delta 
\phantom{0} \gamma \phantom{0}}(x/y)}
\end{equation}
\begin{equation}
\label{16}
\footnotesize{
{\phantom{0}}_{\sigma s}R^{\alpha \phantom{0} i
\phantom{0}}_{\alpha^\prime i^\prime}(y/x)
{\phantom{0}}_{s s}{\cal U}^{i^\prime}_{j^\prime}(x)
{\phantom{0}}_{\sigma s}{R_-}^{\alpha^\prime
j^\prime}_{\beta^\prime \phantom{0}j \phantom{0}}
{\phantom{0}}_{\sigma s}{\cal U}^{\beta^\prime}_{\beta}(y) =
{\phantom{0}}_{\sigma s}{\cal U}^{\alpha}_{\alpha^\prime}(y)
{\phantom{0}}_{\sigma s}{R^+}^{\alpha^\prime \phantom{0}
i}_{\beta^\prime i^\prime}
{\phantom{0}}_{s s}{\cal U}^{i^\prime}_{j^\prime}(x)
{\phantom{0}}_{\sigma s}(R^{-1})^{\beta^\prime
j^\prime}_{\beta \phantom{0} j
\phantom{0}}(x/y)}  \, \, \, 
\end{equation}
\begin{equation}
\label{extra}
\footnotesize{
{\phantom{0}}_{s s}R^{\alpha \phantom{0} i
\phantom{0}}_{\alpha^\prime i^\prime}(y/x)
{\phantom{0}}_{s s}{\cal U}^{i^\prime}_{j^\prime}(x)
{\phantom{0}}_{s s}{R_-}^{\alpha^\prime
j^\prime}_{\beta^\prime \phantom{0}j \phantom{0}}
{\phantom{0}}_{s s}{\cal U}^{\beta^\prime}_{\beta}(y) =
{\phantom{0}}_{s s}{\cal U}^{\alpha}_{\alpha^\prime}(y)
{\phantom{0}}_{s s}{R^+}^{\alpha^\prime \phantom{0}
i}_{\beta^\prime i^\prime}
{\phantom{0}}_{s s}{\cal U}^{i^\prime}_{j^\prime}(x)
{\phantom{0}}_{s s}(R^{-1})^{\beta^\prime
j^\prime}_{\beta \phantom{0} j
\phantom{0}}(x/y)}  \, \, \, .
\end{equation} 
Above we have defined (for $R$-matrices acting on any two spaces)
$$R^+=\lim_{x\rightarrow \infty} R(x).$$ 
For later use we also define 
$$R^-=\lim_{x\rightarrow 0} R(x). $$ 

Equation (\ref{16}) is depicted graphically below. Similar graphical
representations apply for eqs. (\ref{15},\ref{extra}) but will not be
presented.
\[
 \unitlength=0.50mm
 \begin{picture}(170.,85.)
 \put(85.,45.){\makebox(0,0)[cc]{$=$}}
 \put(5.,10.){\vector(0,1){75.}}
 \put(36.,10.){\vector(0,1){75.}}
 \put(15.,10.){\vector(0,1){75.}}
 \put(115.,10.){\vector(0,1){75.}}
 \put(146.,10.){\vector(0,1){75.}}
 \put(125.,10.){\vector(0,1){75.}}
 \put(59.,80.){\vector(1,0){1.}}
 \put(65.,80.){\makebox(0,0)[cc]{$\s \alpha$}}
 \put(65.,70.){\makebox(0,0)[cc]{$\s i $}}
 \put(65.,50.){\makebox(0,0)[cc]{$\s j$}}
 \put(65.,20.){\makebox(0,0)[cc]{$\s \beta $}}
 \put(170.,70.){\makebox(0,0)[cc]{$\s \alpha$}}
 \put(170.,40.){\makebox(0,0)[cc]{$\s i $}}
 \put(170.,20.){\makebox(0,0)[cc]{$\s j$}}
 \put(170.,10.){\makebox(0,0)[cc]{$\s \beta $}}

 \put(-10.,60.){\line(1,1){10.}}
 \put(-10.,60.){\line(1,-1){10.}}
 \put(-10.00,60.00){\circle*{2.}}
 \put(0.,70.){\line(1,0){3.}}
 \put(8.,70.){\line(1,0){4.}}
 \put(17.,70.){\line(1,0){3.5}}
 \put(30.,70.){\line(1,0){3.5}}
 \put(38.,70.){\vector(1,0){22.}}
 \put(25.,70.){\makebox(0,0)[cc]{$\s \ldots$}}

 \put(20.,50.){\vector(-1,0){20.}}
 \put(30.,50.){\line(1,0){30}}
 \put(110.,40.){\line(1,0){3.}}
 \put(118.,40.){\line(1,0){4.}}
 \put(127.,40.){\line(1,0){3.5}}
 \put(144.,40.){\line(-1,0){3.5}}
 \put(148.,40.){\line(1,0){3.5}}
\put(157.,40.){\vector(1,0){10.}}
 \put(110.,20.){\line(1,0){19.}}
 \put(141.,20.){\line(1,0){26.}}
 \put(112.,20.){\vector(-1,0){2.}}
 \put(25.,50.){\makebox(0,0)[cc]{$\s \ldots$}}
 \put(136.,40.){\makebox(0,0)[cc]{$\s \ldots$}}
 \put(136.,20.){\makebox(0,0)[cc]{$\s \ldots$}}

 \put(100.,30.){\line(1,1){10.}}
 \put(100.,30.){\line(1,-1){10.}}
 \put(100.00,30.00){\circle*{2.}}

\put(-10.00,30.00){\circle*{2.3}}
\put(-9.00,31.00){\circle{2.}}
\put(-8.00,32.00){\circle{2.}}
\put(-7.00,33.00){\circle{2.}}
\put(-6.00,34.00){\circle{2.}}
\put(-5.00,35.00){\circle{2.}}
\put(-4.00,36.00){\circle{2.}}
\put(-3.00,37.00){\circle{2.}}
\put(-2.00,38.00){\circle{2.}}
\put(-1.00,39.00){\circle{2.}}
\put(0.00,40.00){\circle{2.}}

\put(-10.00,30.00){\circle{2.}}
\put(-9.00,29.00){\circle{2.}}
\put(-8.00,28.00){\circle{2.}}
\put(-7.00,27.00){\circle{2.}}
\put(-6.00,26.00){\circle{2.}}
\put(-5.00,25.00){\circle{2.}}
\put(-4.00,24.00){\circle{2.}}
\put(-3.00,23.00){\circle{2.}}
\put(-2.00,22.00){\circle{2.}}
\put(-1.00,21.00){\circle{2.}}
\put(0.00,20.00){\circle{2.}}

\put(100.00,60.00){\circle*{2.3}}
\put(101.00,61.00){\circle{2.}}
\put(102.00,62.00){\circle{2.}}
\put(103.00,63.00){\circle{2.}}
\put(104.00,64.00){\circle{2.}}
\put(105.00,65.00){\circle{2.}}
\put(106.00,66.00){\circle{2.}}
\put(107.00,67.00){\circle{2.}}
\put(108.00,68.00){\circle{2.}}
\put(109.00,69.00){\circle{2.}}
\put(110.00,70.00){\circle{2.}}

\put(100.00,60.00){\circle{2.}}
\put(101.00,59.00){\circle{2.}}
\put(102.00,58.00){\circle{2.}}
\put(103.00,57.00){\circle{2.}}
\put(104.00,56.00){\circle{2.}}
\put(105.00,55.00){\circle{2.}}
\put(106.00,54.00){\circle{2.}}
\put(107.00,53.00){\circle{2.}}
\put(108.00,52.00){\circle{2.}}
\put(109.00,51.00){\circle{2.}}
\put(110.00,50.00){\circle{2.}}

\put(111.00,50.00){\circle{2.}}
\put(112.00,50.00){\circle{2.}}
\put(113.00,50.00){\circle{2.}}
\put(114.00,50.00){\circle{2.}}
\put(115.00,50.00){\circle{2.}}
\put(116.00,50.00){\circle{2.}}
\put(117.00,50.00){\circle{2.}}
\put(118.00,50.00){\circle{2.}}
\put(119.00,50.00){\circle{2.}}
\put(120.00,50.00){\circle{2.}}
\put(121.00,50.00){\circle{2.}}
\put(122.00,50.00){\circle{2.}}
\put(123.00,50.00){\circle{2.}}
\put(124.00,50.00){\circle{2.}}
\put(125.00,50.00){\circle{2.}}
\put(126.00,50.00){\circle{2.}}
\put(127.00,50.00){\circle{2.}}
\put(128.00,50.00){\circle{2.}}
\put(129.00,50.00){\circle{2.}}
\put(130.00,50.00){\circle{2.}}
\put(131.00,50.00){\circle{2.}}
\put(136.,50.){\makebox(0,0)[cc]{$\s \ldots$}}
\put(142.00,50.00){\circle{2.}}
\put(143.00,50.00){\circle{2.}}
\put(144.00,50.00){\circle{2.}}
\put(145.00,50.00){\circle{2.}}
\put(146.00,50.00){\circle{2.}}
\put(147.00,50.00){\circle{2.}}
\put(148.00,50.00){\circle{2.}}
\put(149.00,50.00){\circle{2.}}
\put(150.00,50.00){\circle{2.}}
\put(151.00,50.00){\circle{2.}}
\put(152.00,50.00){\circle{2.}}
\put(153.00,50.00){\circle{2.}}
\put(154.00,50.00){\circle{2.}}
\put(154.00,49.00){\circle{2.}}
\put(154.00,48.00){\circle{2.}}
\put(154.00,47.00){\circle{2.}}
\put(154.00,46.00){\circle{2.}}
\put(154.00,45.00){\circle{2.}}
\put(154.00,44.00){\circle{2.}}
\put(154.00,43.00){\circle{2.}}
\put(154.00,42.00){\circle{2.}}
\put(154.00,41.00){\circle{2.}}
\put(154.00,40.00){\circle{2.}}
\put(154.00,39.00){\circle{2.}}
\put(154.00,38.00){\circle{2.}}
\put(154.00,37.00){\circle{2.}}
\put(154.00,36.00){\circle{2.}}
\put(154.00,35.00){\circle{2.}}
\put(154.00,34.00){\circle{2.}}
\put(154.00,33.00){\circle{2.}}
\put(154.00,32.00){\circle{2.}}
\put(154.00,31.00){\circle{2.}}
\put(154.00,30.00){\circle{2.}}
\put(154.00,29.00){\circle{2.}}
\put(154.00,28.00){\circle{2.}}
\put(154.00,27.00){\circle{2.}}
\put(154.00,26.00){\circle{2.}}
\put(154.00,25.00){\circle{2.}}
\put(154.00,24.00){\circle{2.}}
\put(154.00,23.00){\circle{2.}}
\put(154.00,22.00){\circle{2.}}
\put(154.00,21.00){\circle{2.}}
\put(154.00,20.00){\circle{2.}}
\put(154.00,19.00){\circle{2.}}
\put(154.00,18.00){\circle{2.}}
\put(154.00,17.00){\circle{2.}}
\put(154.00,16.00){\circle{2.}}
\put(154.00,15.00){\circle{2.}}
\put(154.00,14.00){\circle{2.}}
\put(154.00,13.00){\circle{2.}}
\put(154.00,12.00){\circle{2.}}
\put(154.00,11.00){\circle{2.}}
\put(154.00,10.00){\circle{2.}}
\put(155.00,10.00){\circle{2.}}
\put(156.00,10.00){\circle{2.}}
\put(157.00,10.00){\circle{2.}}
\put(158.00,10.00){\circle{2.}}
\put(159.00,10.00){\circle{2.}}
\put(160.00,10.00){\circle{2.}}
\put(161.00,10.00){\circle{2.}}
\put(162.00,10.00){\circle{2.}}
\put(163.00,10.00){\circle{2.}}
\put(164.00,10.00){\circle{2.}}
\put(165.00,10.00){\circle{2.}}
\put(166.00,10.00){\circle{2.}}

\put(1.00,20.00){\circle{2.}}
\put(2.00,20.00){\circle{2.}}
\put(3.00,20.00){\circle{2.}}
\put(4.00,20.00){\circle{2.}}
\put(5.00,20.00){\circle{2.}}
\put(6.00,20.00){\circle{2.}}
\put(7.00,20.00){\circle{2.}}
\put(8.00,20.00){\circle{2.}}
\put(9.00,20.00){\circle{2.}}
\put(10.00,20.00){\circle{2.}}
\put(11.00,20.00){\circle{2.}}
\put(12.00,20.00){\circle{2.}}
\put(13.00,20.00){\circle{2.}}
\put(14.00,20.00){\circle{2.}}
\put(15.00,20.00){\circle{2.}}
\put(16.00,20.00){\circle{2.}}
\put(17.00,20.00){\circle{2.}}
\put(18.00,20.00){\circle{2.}}
\put(19.00,20.00){\circle{2.}}
\put(25.,20.){\makebox(0,0)[cc]{$\s \ldots$}}
\put(31.00,20.00){\circle{2.}}
\put(32.00,20.00){\circle{2.}}
\put(33.00,20.00){\circle{2.}}
\put(34.00,20.00){\circle{2.}}
\put(35.00,20.00){\circle{2.}}
\put(36.00,20.00){\circle{2.}}
\put(37.00,20.00){\circle{2.}}
\put(38.00,20.00){\circle{2.}}
\put(39.00,20.00){\circle{2.}}
\put(40.00,20.00){\circle{2.}}
\put(41.00,20.00){\circle{2.}}
\put(42.00,20.00){\circle{2.}}
\put(43.00,20.00){\circle{2.}}
\put(44.00,20.00){\circle{2.}}
\put(45.00,20.00){\circle{2.}}
\put(46.00,20.00){\circle{2.}}
\put(47.00,20.00){\circle{2.}}
\put(48.00,20.00){\circle{2.}}
\put(49.00,20.00){\circle{2.}}
\put(50.00,20.00){\circle{2.}}
\put(51.00,20.00){\circle{2.}}
\put(52.00,20.00){\circle{2.}}
\put(53.00,20.00){\circle{2.}}
\put(54.00,20.00){\circle{2.}}
\put(55.00,20.00){\circle{2.}}
\put(56.00,20.00){\circle{2.}}
\put(57.00,20.00){\circle{2.}}
\put(58.00,20.00){\circle{2.}}
\put(59.00,20.00){\circle{2.}}
\put(60.00,20.00){\circle{2.}}
\put(61.00,20.00){\circle{2.}}

\put(1.00,40.00){\circle{2.}}
\put(2.00,40.00){\circle{2.}}
\put(3.00,40.00){\circle{2.}}
\put(7.00,40.00){\circle{2.}}
\put(8.00,40.00){\circle{2.}}
\put(9.00,40.00){\circle{2.}}
\put(10.00,40.00){\circle{2.}}
\put(12.,40.){\vector(1,0){1.}}
\put(17.00,40.00){\circle{2.}}
\put(18.00,40.00){\circle{2.}}
\put(19.00,40.00){\circle{2.}}
\put(21.,40.){\vector(1,0){1.}}
\put(25.,40.){\makebox(0,0)[cc]{$\s \ldots$}}
\put(30.00,40.00){\circle{2.}}
\put(31.00,40.00){\circle{2.}}
\put(32.00,40.00){\circle{2.}}
\put(33.00,40.00){\circle{2.}}
\put(34.00,40.00){\circle{2.}}
\put(39.00,40.00){\circle{2.}}
\put(40.00,40.00){\circle{2.}}
\put(41.00,40.00){\circle{2.}}
\put(42.00,40.00){\circle{2.}}
\put(43.00,40.00){\circle{2.}}
\put(44.00,40.00){\circle{2.}}
\put(45.00,40.00){\circle{2.}}
\put(45.00,41.00){\circle{2.}}
\put(45.00,42.00){\circle{2.}}
\put(45.00,43.00){\circle{2.}}
\put(45.00,44.00){\circle{2.}}
\put(45.00,45.00){\circle{2.}}
\put(45.00,46.00){\circle{2.}}
\put(45.00,47.00){\circle{2.}}
\put(45.00,52.00){\circle{2.}}
\put(45.00,53.00){\circle{2.}}
\put(45.00,54.00){\circle{2.}}
\put(45.00,55.00){\circle{2.}}
\put(45.00,56.00){\circle{2.}}
\put(45.00,57.00){\circle{2.}}
\put(45.00,58.00){\circle{2.}}
\put(45.00,59.00){\circle{2.}}
\put(45.00,60.00){\circle{2.}}
\put(45.00,61.00){\circle{2.}}
\put(45.00,62.00){\circle{2.}}
\put(45.00,63.00){\circle{2.}}
\put(45.00,64.00){\circle{2.}}
\put(45.00,65.00){\circle{2.}}
\put(45.00,66.00){\circle{2.}}
\put(45.00,67.00){\circle{2.}}
\put(45.00,68.00){\circle{2.}}
\put(45.00,69.00){\circle{2.}}
\put(45.00,70.00){\circle{2.}}
\put(45.00,71.00){\circle{2.}}
\put(45.00,72.00){\circle{2.}}
\put(45.00,73.00){\circle{2.}}
\put(45.00,74.00){\circle{2.}}
\put(45.00,75.00){\circle{2.}}
\put(45.00,76.00){\circle{2.}}
\put(45.00,77.00){\circle{2.}}
\put(45.00,78.00){\circle{2.}}
\put(45.00,79.00){\circle{2.}}
\put(45.00,80.00){\circle{2.}}
\put(45.00,80.00){\circle{2.}}
\put(46.00,80.00){\circle{2.}}
\put(47.00,80.00){\circle{2.}}
\put(48.00,80.00){\circle{2.}}
\put(49.00,80.00){\circle{2.}}
\put(50.00,80.00){\circle{2.}}
\put(51.00,80.00){\circle{2.}}
\put(52.00,80.00){\circle{2.}}
\put(53.00,80.00){\circle{2.}}
\put(54.00,80.00){\circle{2.}}
\put(55.00,80.00){\circle{2.}}
\put(56.00,80.00){\circle{2.}}
\put(57.00,80.00){\circle{2.}}

\put(25.,20.){\makebox(0,0)[cc]{$\s \ldots$}}
\put(33.00,20.00){\circle{2.}}
\put(34.00,20.00){\circle{2.}}
\put(35.00,20.00){\circle{2.}}
\put(36.00,20.00){\circle{2.}}
\put(37.00,20.00){\circle{2.}}
\put(38.00,20.00){\circle{2.}}
\put(39.00,20.00){\circle{2.}}
\put(40.00,20.00){\circle{2.}}
\put(41.00,20.00){\circle{2.}}
\put(42.00,20.00){\circle{2.}}
\put(43.00,20.00){\circle{2.}}
\put(44.00,20.00){\circle{2.}}
\put(45.00,20.00){\circle{2.}}
\put(46.00,20.00){\circle{2.}}
\put(47.00,20.00){\circle{2.}}
\put(48.00,20.00){\circle{2.}}
\put(49.00,20.00){\circle{2.}}
\put(50.00,20.00){\circle{2.}}
\put(51.00,20.00){\circle{2.}}
\put(52.00,20.00){\circle{2.}}
\put(53.00,20.00){\circle{2.}}
\put(54.00,20.00){\circle{2.}}
\put(55.00,20.00){\circle{2.}}
\put(56.00,20.00){\circle{2.}}
\put(57.00,20.00){\circle{2.}}
\put(58.00,20.00){\circle{2.}}
\put(59.00,20.00){\circle{2.}}
\put(60.00,20.00){\circle{2.}}
\put(61.00,20.00){\circle{2.}}
\put(112.00,70.00){\circle{2.}}
\put(113.00,70.00){\circle{2.}}
\put(117.00,70.00){\circle{2.}}
\put(118.00,70.00){\circle{2.}}
\put(119.00,70.00){\circle{2.}}
\put(120.00,70.00){\circle{2.}}
\put(122.,70.){\vector(1,0){1.}}
\put(129.00,70.00){\circle{2.}}
\put(130.00,70.00){\circle{2.}}
\put(127.00,70.00){\circle{2.}}
\put(128.00,70.00){\circle{2.}}
\put(132.,70.){\vector(1,0){1.}}
\put(136.,70.){\makebox(0,0)[cc]{$\s \ldots$}}
\put(142.00,70.00){\circle{2.}}
\put(143.00,70.00){\circle{2.}}
\put(144.00,70.00){\circle{2.}}
\put(148.00,70.00){\circle{2.}}
\put(149.00,70.00){\circle{2.}}
\put(150.00,70.00){\circle{2.}}
\put(151.00,70.00){\circle{2.}}
\put(152.00,70.00){\circle{2.}}
\put(153.00,70.00){\circle{2.}}
\put(154.00,70.00){\circle{2.}}
\put(155.00,70.00){\circle{2.}}
\put(156.00,70.00){\circle{2.}}
\put(157.00,70.00){\circle{2.}}
\put(158.00,70.00){\circle{2.}}
\put(159.00,70.00){\circle{2.}}
\put(160.00,70.00){\circle{2.}}
\put(161.00,70.00){\circle{2.}}
\put(162.00,70.00){\circle{2.}}
\put(163.00,70.00){\circle{2.}}
\put(164.00,70.00){\circle{2.}}
\put(166.,70.){\vector(1,0){1.}}
 \end{picture}
 \]

Finally, the spin-1 transfer matrix
is constructed by taking the spin-1 Markov trace
of the monodromy matrix (\ref{10})
in the auxiliary space
\begin{equation}
\label{17}
\phantom{0}_{ss}{\cal T}^{ \{ j \} }_{ \, \{i\} }(x) =
   \sum_{\alpha} \phantom{0}_{s}K^\alpha_\alpha
\phantom{0}_{ss}{\cal{U}}^{\alpha \{ j \} }_{
\alpha \{ i \} } \, \, \,
=    \, \, \,
\ba{c}
\unitlength=0.50mm
\begin{picture}(95.,49.)
\put(46.,15.){\makebox(0,0)[cc]{$\cdots$}}
\put(21.,-3.){\makebox(0,0)[cc]{$\s i_1$}}
\put(31.,-3.){\makebox(0,0)[cc]{$\s i_2$}}
\put(71.,-3.){\makebox(0,0)[cc]{$\s i_L$}}
\put(21.,47.){\makebox(0,0)[cc]{$\s j_1$}}
\put(31.,47.){\makebox(0,0)[cc]{$\s j_2$}}
\put(71.,47.){\makebox(0,0)[cc]{$\s j_L$}}
\put(70.,0.){\vector(0,1){45.}}
\put(30.,0.){\vector(0,1){45.}}
\put(20.,0.){\vector(0,1){45.}}
\put(73.,35.){\vector(1,0){8.}}
\put(57.,15.){\line(1,0){25.}}
\put(47.,35.){\makebox(0,0)[cc]{$\cdots$}}
\put(15.,15.){\line(1,0){21.}}
\put(30.,15.){\vector(-1,0){15.}}
\put(22.,35.){\line(1,0){6.}}
\put(32.,35.){\line(1,0){6.}}
\put(57.,35.){\line(1,0){8.}}
\put(80.,35.){\line(0,0){0.}}
\put(15.,35.){\line(1,0){4.}}

\put(5.,25.){\line(1,1){10.}}
\put(5.,25.){\line(1,-1){10.}}
\put(5.,25.00){\circle*{2.}}

\put(82.,15.){\line(1,-1){7.1}}
\put(81.,35.){\line(1,1){7.85}}

\put(89.3,8.0){\line(1,1){17.3}}
\put(89.3,43.){\line(1,-1){17.3}}
\end{picture}
\ea\
\end{equation}
where
\begin{equation}
\label{18}
\phantom{0}_{s} K \, = \, \pmatrix{ q^2 \, \phantom{000} \, \phantom{000} \cr
          \phantom{000} \, 1 \, \phantom{000} \cr
          \phantom{000} \, \phantom{000} \, q^{-2} \cr}  \, \, \, ,
\end{equation}
By using equations (\ref{cross},\ref{extra}) it can be shown
that this transfer matrix forms a commuting family,
i.e., it commutes for different spectral parameters.
A quantum algebra invariant spin-1 XXZ Hamiltonian with closed boundary
conditions will be obtained later from it. However,
in order to diagonalize it, the usual algebraic Bethe
ansatz scheme
which applies to monodromy matrices whose auxiliary space
is the fundamental representation can not be adopted. As in other
higher-spin chains \cite{Mnr,1,2,bab1,3} this problem can
be solved by introducing  an auxiliary spin-1/2
transfer matrix $ \phantom{0}_{\sigma s}{\cal T}$
which commutes with the spin-1 transfer matrix
$ \phantom{0}_{ss}{\cal T}$.
This spin-1/2 auxiliary transfer matrix is constructed
using the auxiliary $ \phantom{0}_{\sigma s} R(x)  $
(\ref{5}) and doubled monodromy $ \phantom{0}_{\sigma s}{\cal U}(x)$
(\ref{13}) matrices and is given by
\begin{equation}
\label{19}
\phantom{0}_{\sigma s}{\cal T}^{ \{ j \} }_{ \{i\} }(x) =
   \sum_{\alpha} \phantom{0}_{\sigma }K^\alpha_\alpha
\phantom{0}_{\sigma s}{\cal{U}}^{\alpha \{ j \} }_{
\alpha \{ i \} } \, \, \,
=    \, \, \,
\ba{c}
\unitlength=0.50mm
\begin{picture}(95.,49.)
\put(45.,15.){\makebox(0,0)[cc]{$\cdots$}}
\put(21.,-3.){\makebox(0,0)[cc]{$\s i_1$}}
\put(31.,-3.){\makebox(0,0)[cc]{$\s i_2$}}
\put(71.,-3.){\makebox(0,0)[cc]{$\s i_L$}}
\put(21.,47.){\makebox(0,0)[cc]{$\s j_1$}}
\put(31.,47.){\makebox(0,0)[cc]{$\s j_2$}}
\put(71.,47.){\makebox(0,0)[cc]{$\s j_L$}}
\put(70.,0.){\vector(0,1){45.}}
\put(30.,0.){\vector(0,1){45.}}
\put(20.,0.){\vector(0,1){45.}}
\put(80.,35.){\vector(1,0){1.}}

\put(57.00,35.00){\circle{2.}}
\put(58.00,35.00){\circle{2.}}
\put(59.00,35.00){\circle{2.}}
\put(60.00,35.00){\circle{2.}}
\put(61.00,35.00){\circle{2.}}
\put(62.00,35.00){\circle{2.}}
\put(63.00,35.00){\circle{2.}}
\put(64.00,35.00){\circle{2.}}
\put(65.00,35.00){\circle{2.}}
\put(66.00,35.00){\circle{2.}}
\put(67.00,35.00){\circle{2.}}
\put(68.00,35.00){\circle{2.}}
\put(72.00,35.00){\circle{2.}}
\put(73.00,35.00){\circle{2.}}
\put(74.00,35.00){\circle{2.}}
\put(75.00,35.00){\circle{2.}}
\put(76.00,35.00){\circle{2.}}
\put(77.00,35.00){\circle{2.}}
\put(78.00,35.00){\circle{2.}}
\put(57.00,15.00){\circle{2.}}
\put(58.00,15.00){\circle{2.}}
\put(59.00,15.00){\circle{2.}}
\put(60.00,15.00){\circle{2.}}
\put(61.00,15.00){\circle{2.}}
\put(62.00,15.00){\circle{2.}}
\put(63.00,15.00){\circle{2.}}
\put(64.00,15.00){\circle{2.}}
\put(65.00,15.00){\circle{2.}}
\put(66.00,15.00){\circle{2.}}
\put(67.00,15.00){\circle{2.}}
\put(68.00,15.00){\circle{2.}}
\put(69.00,15.00){\circle{2.}}
\put(70.00,15.00){\circle{2.}}
\put(71.00,15.00){\circle{2.}}
\put(72.00,15.00){\circle{2.}}
\put(73.00,15.00){\circle{2.}}
\put(74.00,15.00){\circle{2.}}
\put(75.00,15.00){\circle{2.}}
\put(76.00,15.00){\circle{2.}}
\put(77.00,15.00){\circle{2.}}
\put(78.00,15.00){\circle{2.}}
\put(79.00,15.00){\circle{2.}}
\put(80.00,15.00){\circle{2.}}
\put(45.,35.){\makebox(0,0)[cc]{$\cdots$}}
\put(15.00,15.00){\circle{2.}}
\put(16.00,15.00){\circle{2.}}
\put(17.00,15.00){\circle{2.}}
\put(18.00,15.00){\circle{2.}}
\put(19.00,15.00){\circle{2.}}
\put(20.00,15.00){\circle{2.}}
\put(21.00,15.00){\circle{2.}}
\put(22.00,15.00){\circle{2.}}
\put(23.00,15.00){\circle{2.}}
\put(24.00,15.00){\circle{2.}}
\put(25.00,15.00){\circle{2.}}
\put(26.00,15.00){\circle{2.}}
\put(27.00,15.00){\circle{2.}}
\put(28.00,15.00){\circle{2.}}
\put(29.00,15.00){\circle{2.}}
\put(30.00,15.00){\circle{2.}}
\put(31.00,15.00){\circle{2.}}
\put(32.00,15.00){\circle{2.}}
\put(33.00,15.00){\circle{2.}}
\put(34.00,15.00){\circle{2.}}
\put(35.00,15.00){\circle{2.}}
\put(36.00,15.00){\circle{2.}}
\put(37.00,15.00){\circle{2.}}
\put(15.00,35.00){\circle{2.}}
\put(16.00,35.00){\circle{2.}}
\put(17.00,35.00){\circle{2.}}
\put(18.00,35.00){\circle{2.}}
\put(22.00,35.00){\circle{2.}}
\put(23.00,35.00){\circle{2.}}
\put(24.00,35.00){\circle{2.}}
\put(25.00,35.00){\circle{2.}}
\put(26.00,35.00){\circle{2.}}
\put(28.00,35.){\vector(1,0){1.}}

\put(32.00,35.00){\circle{2.}}
\put(33.00,35.00){\circle{2.}}
\put(34.00,35.00){\circle{2.}}
\put(35.00,35.00){\circle{2.}}
\put(36.00,35.00){\circle{2.}}
\put(37.00,35.00){\circle{2.}}
\put(39.00,35.){\vector(1,0){1.}}

\put(5.00,25.){\circle*{2.5}}

\put(5.00,25.00){\circle{2.}}
\put(6.00,26.00){\circle{2.}}
\put(7.00,27.00){\circle{2.}}
\put(8.00,28.00){\circle{2.}}
\put(9.00,29.00){\circle{2.}}
\put(10.00,30.00){\circle{2.}}
\put(11.00,31.00){\circle{2.}}
\put(12.00,32.00){\circle{2.}}
\put(13.00,33.00){\circle{2.}}
\put(14.00,34.00){\circle{2.}}
\put(15.00,35.00){\circle{2.}}

\put(5.00,25.00){\circle{2.}}
\put(6.00,24.00){\circle{2.}}
\put(7.00,23.00){\circle{2.}}
\put(8.00,22.00){\circle{2.}}
\put(9.00,21.00){\circle{2.}}
\put(10.00,20.00){\circle{2.}}
\put(11.00,19.00){\circle{2.}}
\put(12.00,18.00){\circle{2.}}
\put(13.00,17.00){\circle{2.}}
\put(14.00,16.00){\circle{2.}}
\put(15.00,15.00){\circle{2.}}

\put(85.00,10.00){\circle{2.}}
\put(84.00,11.00){\circle{2.}}
\put(83.00,12.00){\circle{2.}}
\put(82.00,13.00){\circle{2.}}
\put(81.00,14.00){\circle{2.}}
\put(80.00,15.00){\circle{2.}}

\put(85.00,40.00){\circle{2.}}
\put(84.00,39.00){\circle{2.}}
\put(83.00,38.00){\circle{2.}}
\put(82.00,37.00){\circle{2.}}
\put(81.00,36.00){\circle{2.}}
\put(80.00,35.00){\circle{2.}}

\put(85.00,40.00){\circle{2.}}
\put(86.00,39.00){\circle{2.}}
\put(87.00,38.00){\circle{2.}}
\put(88.00,37.00){\circle{2.}}
\put(89.00,36.00){\circle{2.}}
\put(90.00,35.00){\circle{2.}}
\put(91.00,34.00){\circle{2.}}
\put(92.00,33.00){\circle{2.}}
\put(93.00,32.00){\circle{2.}}
\put(94.00,31.00){\circle{2.}}
\put(95.00,30.00){\circle{2.}}
\put(96.00,29.00){\circle{2.}}
\put(97.00,28.00){\circle{2.}}
\put(98.00,27.00){\circle{2.}}
\put(99.00,26.00){\circle{2.}}
\put(100.00,25.00){\circle{2.}}

\put(85.00,10.00){\circle{2.}}
\put(86.00,11.00){\circle{2.}}
\put(87.00,12.00){\circle{2.}}
\put(88.00,13.00){\circle{2.}}
\put(89.00,14.00){\circle{2.}}
\put(90.00,15.00){\circle{2.}}
\put(91.00,16.00){\circle{2.}}
\put(92.00,17.00){\circle{2.}}
\put(93.00,18.00){\circle{2.}}
\put(94.00,19.00){\circle{2.}}
\put(95.00,20.00){\circle{2.}}
\put(96.00,21.00){\circle{2.}}
\put(97.00,22.00){\circle{2.}}
\put(98.00,23.00){\circle{2.}}
\put(99.00,24.00){\circle{2.}}
\put(100.00,25.00){\circle{2.}}

\end{picture}
\ea\ ,
\end{equation}
with
\begin{equation}
\label{20}
\phantom{0}_{\sigma } K \, = \, \pmatrix{ q \,  \phantom{000} \cr
             \phantom{000} \,  q^{-1} \cr}  \, \, \, ,
\end{equation}
Using eqs. (\ref{cross},\ref{16})  we can also show that the above transfer matrices commute
\begin{equation}
[ \phantom{0}_{\sigma s} {\cal T} (x) \, , \,
\phantom{0}_{s s} {\cal T} (y) \, ] \, = \, 0 \, \, \, \, .
\label{21}
\end{equation}
Therefore we can simultaneously diagonalize
$ \phantom{0}_{\sigma s}{\cal T} $ and $ \phantom{0}_{s s} {\cal T}$
which will be presented in the next section.

A quantum algebra invariant closed spin-1 Hamiltonian can be defined through
\begin{equation}
\label{ham}
{\cal{H}} \, = \, \phantom{0}_{s s}{\cal T}^{\prime} (x) \,
               \phantom{0}_{s s}{\cal T}^{-1} (x) \,
               \left|_{x=1}\right. \, \, \, ,
\end{equation}
where the prime indicates differentiation with respect to the
variable $x$. This yields (see \cite{Jonangi} for details about this general
construction)
\begin{equation}
\label{22}
{\cal{H}} \, = \, \sum_{n = 1}^{L-1} h_n + h_0 \, \, \, ,
\end{equation}
where
\begin{eqnarray}
\label{23}
 h_n \, & \propto & \, \vec{J_n} \, .\,  \vec{J}_{n+1} \, - \,
( \vec{J_n} \, . \,  \vec{J}_{n+1} )^2 \, + \,
\frac{ {(q-q^{-1})}^2}{2}
\biggl[ J_n^z J_{n+1}^z \, + \,
(J_{n}^z)^2 \, + \, (J_{n+1}^z)^2 \, - \,
(J_{n}^z J_{n+1}^z)^2 \biggr] \nonumber \\
\, &-& \, ({q^{1/2}-q^{-1/2})^2}
\biggl[ ( J_n^x J_{n+1}^x  +  J_n^y J_{n+1}^y ) J_n^z J_{n+1}^z
\, + \, J_n^z J_{n+1}^z ( J_n^x J_{n+1}^x + J_n^y J_{n+1}^y )
\biggr]
\end{eqnarray}
and $\vec{J_n}$ are spin-1 generators of $sl(2)$. The boundary term
$h_0$ is given by
\begin{equation}
\label{24}
h_0 = \underbrace{\phantom{0}_{s s} {\hat{R}}^{-}_{1}
\phantom{0}_{s s}{\hat{R}}^{-}_{2}
\dots
\phantom{0}_{s s}{\hat{R}}^{-}_{L-1}}_{\mbox{{G}} }
\, h_{L-1} \, \, \,
\underbrace{
\phantom{0}_{s s}{\hat{R}}^{+}_{L-1}
\dots
\phantom{0}_{s s}{\hat{R}}^{+}_{2}
\phantom{0}_{s s}{\hat{R}}^{+}_{1}}_{\mbox{{$G^{-1}$}} } \, \, \, ,
\end{equation}
with
\begin{equation}
\label{25}
\phantom{0}_{s s}{\hat{R}}^{ \pm \, \{ \gamma \} }_{ n \,\, \{ \beta \} } \, = \,
{\bf{1}}^{\gamma_1}_{\beta_1} \otimes
{\bf{1}}^{\gamma_{2}}_{\beta_{2}} \otimes \dots
\otimes {\phantom{0}_{s s}{R}^{\pm}}^{\, \, \,  \gamma_n \gamma_{n+1} }_{ \, \, \, \beta_{n+1}
\beta_n} \otimes
\dots  {\bf{1}}^{\gamma_L}_{\beta_L} \, \, \, \, \, \, \, \, \, \,
\, \, \, \, n = 1, 2,  \dots L-1 \, \, \, .
\end{equation}
In eq. (\ref{22})  $L$ is the number of lattice sites.
The operators $H$, $h_n$ and $ {\hat{R}}^{\pm}_n$ ($n= 1, 2,  \dots
L-1$) act on the
``quantum space'' ${\bf C}^{3L}$ ( for simplicity, we omit the
quantum space indices and write them only whenever necessary).
The model is periodic in the sense that
the operator $G^{-1}$ maps $h_{n}$ into $h_{n-1}$
\begin{equation}
\label{26}
G^{-1} h_{n} G \, = \, h_{n-1 } \, \, \,
\, \, \, \, \, \, \, \, \, \, \,
n = 2, \dots L-1 \, \, \, ,
\end{equation}
and $h_1$ into $h_0$
\begin{equation}
\label{27}
G^{-1} h_{1} G \, = \, G h_{L-1} G^{-1} \, \, \, .
\end{equation}
The quantum algebra invariance of such a construction is discussed in
\cite{Jonangi}.

\section*{III. Bethe ansatz method}
In this section we solve the eigenvalue problem of the
transfer matrix
\begin{equation}
\label{28}
\phantom{0}_{s s} {\cal T} \Psi = (q^2  \phantom{0}_{s s}{\cal{U}}^{1}_{1}
\, + \, \phantom{0}_{s s}{\cal{U}}^{2}_{2} \, + \,
q^{-2} \phantom{0}_{s s}{\cal{U}}^{3}_{3}
) \Psi =
\phantom{0}_{s s}\Lambda \Psi \, \, \, ,
\end{equation}
(and consequently that of the Hamiltonian~(\ref{22})) through a
combination of the techniques developed to handle with
quantum group invariant closed chains \cite{Kar}
and higher-spin chains \cite{Mnr}. First, from the
fact that eq. (\ref{21}) is satisfied, $ \phantom{0}_{s s} {\cal T} $
and $\phantom{0}_{\sigma s} {\cal T}$ have a common set of
eigenvectors, which can be determined by applying the algebraic
Bethe ansatz method to $\phantom{0}_{\sigma s} {\cal T}$ (\ref{19}).
Following Babujian \cite{bab1}, the vector $\Psi$ can be written as
\begin{equation}
\label{29}
\Psi = B(x_1)\, B(x_2)\, \dots B(x_M) \Phi
\end{equation}
where  $ \Phi$ is the reference state defined by the equation
\[
C \Phi = 0
\]
whose solution is $\Phi =\otimes^{L}_{i=1}{|1>}_i $. It is an
eigenstate of $A$  and $D$
\begin{eqnarray}
A(x)\Phi \, &=& \, q^{\frac{3L}2}\, \Phi \, \, \, ,
\nonumber \\
D(x)\Phi &=& q^{-\frac L2} \frac{\phantom{0}_{\sigma s}c(1/x)^L}{ 
\phantom{0}_{\sigma s}a(1/x)^{L}} \Phi \, \, \, .
\label{30}
\end{eqnarray}
Next we apply $A(x)$ and $D(x)$ to $\Psi$ (\ref{29}), push them
through all the $B's$ using the following commutation rules
derived from the Yang-Baxter relation (\ref{15})
\begin{eqnarray}
A(x) B(y) &=&
{ 1 \over q } { { \phantom{0}_{\sigma \sigma} a(x/y)}  \over
{ \phantom{0}_{\sigma \sigma}b(x/y)} }
B(y) A(x) - {1 \over q}
{ {\phantom{0}_{\sigma \sigma} c_- (x/y)} \over
{\phantom{0}_{\sigma \sigma}b(x/y)} } B(x) A(y) -
{ {q-1/q} \over q }
B(x) D(y) \,
\nonumber \\
D(x) B(y) &=&
q {\phantom{0}_{\sigma \sigma}a(y/x) \over \phantom{0}_{\sigma \sigma}b(y/x) }
B(y) D(x) \, - \,
q {\phantom{0}_{\sigma \sigma} c_-(y/x) \over
\phantom{0}_{\sigma \sigma}b(y/x) } B(x) D(y)
\label{31}
\, ,
\end{eqnarray}
and apply them to the reference state $\Phi$ using eqs. (\ref{30}).
>From the first terms of the r.h.s. of eqs. (\ref{30}) we get the
``wanted'' contributions, while the other terms originate the
``unwanted'' terms, since they can never give a vector proportional
to $\Psi$
\begin{eqnarray}
A(x) \Psi &=& q^{\frac{3L}2 - M} 
\prod_{i=1}^{M}
{\phantom{0}_{\sigma \sigma}a(x/x_i) \over
\phantom{0}_{\sigma \sigma}b(x/x_i)}
\Psi \, + \, \mbox{u. t.} \, \nonumber \\
D(x) \Psi &=& q^{-\frac L2 + M} \frac{{\phantom{0}_{\sigma s}c(1/x)}^L
}{{\phantom{0}_{\sigma s}a(1/x)}^L} \prod_{i=1}^{M}
{\phantom{0}_{\sigma \sigma}a(x_i/x) \over
\phantom{0}_{\sigma \sigma}b(x_i/x)}
\Psi \, + \, \mbox{u. t.} \,
\label{32}
\end{eqnarray}
The cancellation of all unwanted terms ensure that $\Psi$, as given
by (\ref{29}) is an eigenstate of the transfer matrix
$\phantom{0}_{\sigma s} {\cal T}(x)$ (\ref{19}) and this
indeed happens if the Bethe ansatz equations (BAE) hold
\begin{equation}
\label{33}
q^{2(1 + L - M)}
{\biggl( {\phantom{0}_{\sigma s}a(1/x_k)
\over \phantom{0}_{\sigma s}c(1/x_k) }  \biggr)}^L
\prod_{i=1}^{M} { \phantom{0}_{\sigma \sigma}a(x_k/x_i)
\over \phantom{0}_{\sigma \sigma}b(x_k/x_i) }
{\phantom{0}_{\sigma \sigma}b(x_i/x_k) \over
\phantom{0}_{\sigma \sigma}a(x_i/x_k) }
= -1 \, \, \, , \, \,
\, \, \, \, \, \, \, \, \, \, \, \,
k = 1, \dots M \, \, \, ,
\end{equation}
Note that these equations are much simpler than those obtained
for the quantum group invariant spin 1 chain with open
boundary conditions (see \cite{Mnr}). Also in the limit
$q \rightarrow 1$ we recover the BAE for the usual
periodic case \cite{bab1}.

Let us now find the eigenvalues of $\phantom{0}_{s s} {\cal T}(x)$
by acting with this transfer matrix on $\Psi$ (\ref{29}),
according to (\ref{28}). For this purpose we need the
commutation relations between $ \phantom{0}_{s s}{\cal{U}}^{1}_{1}(x)$,
$ \phantom{0}_{s s}{\cal{U}}^{2}_{2}(x), \phantom{0}_{s s}
{\cal{U}}^{3}_{3}(x) $ and $ \phantom{0}_{s s}{\cal{U}}^{1}_{2}(y) $
and their action on the reference state $\Phi$. 
Rewriting the Yang-Baxter equation (\ref{16}) 
we can find the following relations
\begin{equation}
\label{37}
\footnotesize{
{\phantom{0}}_{\sigma s}{\cal U}^{\alpha}_{\alpha^\prime}(y)
{\phantom{0}}_{\sigma s}{R^+}^{\alpha^\prime \phantom{0}
i}_{\beta \phantom{0} i^\prime \phantom{0}}
{\phantom{0}}_{s s}{\cal U}^{i^\prime}_{j}(x) =
{\phantom{0}}_{\sigma s}R^{\alpha \phantom{0}
i}_{\alpha^\prime i^\prime}(y/x)
{\phantom{0}}_{s s}{\cal U}^{i^\prime}_{j^\prime}(x)
{\phantom{0}}_{\sigma s}{R_-}^{\alpha^\prime \phantom{0}
j^\prime}_{\beta^\prime j^{ \prime \prime}}
{\phantom{0}}_{\sigma s}{\cal U}^{\beta^\prime}_{\beta^{\prime \prime}}(y)
{\phantom{0}}_{\sigma s} R^{\beta^{\prime \prime} j^{\prime \prime}}_{
\beta \phantom{0} j\phantom{0}}(x/y)}
\end{equation}
which yield the commutation rules
\begin{eqnarray}
\phantom{0}_{s s}{\cal{U}}^{1}_{1}(x) B(y) &=&
{ 1 \over q^2}{\phantom{0}_{\sigma s}a(x/y) \over {
\phantom{0}_{\sigma s}c(x/y)} } B(y)
\phantom{0}_{s s}{\cal{U}}^{1}_{1}(x)
-{1 \over q}{\phantom{0}_{\sigma s}d_-(x/y) \over
\phantom{0}_{\sigma s}c(x/y) }
\phantom{0}_{s s}{\cal{U}}^{1}_{2}(x) A(y)
\nonumber \\
&-& {1 \over q} \sqrt{ {(1 - {1 \over q^2})(q^2 - {1\over {q^2}})} }
\biggl( \phantom{0}_{s s}{\cal{U}}^{1}_{2}(x) D(y)
+ {\phantom{0}_{\sigma s}d_-(x/y) \over
\phantom{0}_{\sigma s}c(x/y) }
\phantom{0}_{s s}{\cal{U}}^{1}_{3}(x) C(y) \biggr)
\nonumber \\
\phantom{0}_{s s}{\cal{U}}^{2}_{2}(x) B(y) &=&
{\phantom{0}_{\sigma s}a(y/x)\phantom{0}_{\sigma s}a(x/y) \over {\phantom{0}_{\sigma s}b(y/x)
\phantom{0}_{\sigma s}b(x/y)} } B(y)
\phantom{0}_{s s}{\cal{U}}^{2}_{2}(x)
-{\phantom{0}_{\sigma s}d_-(y/x) \over
\phantom{0}_{\sigma s}b(y/x) }
\phantom{0}_{s s}{\cal{U}}^{1}_{2}(x) D(y)
\nonumber \\
&-& {\phantom{0}_{\sigma s}d_-(x/y) \over
\phantom{0}_{\sigma s}b(x/y) }
\biggl( {1 \over q} \phantom{0}_{s s}{\cal{U}}^{2}_{3}(x) A(y)
+ q { {\phantom{0}_{\sigma s}d_-(y/x)} \over
{\phantom{0}_{\sigma s}b(y/x)} }
\phantom{0}_{s s}{\cal{U}}^{1}_{3}(x) C(y) \biggr)
\nonumber \\
&-&   \sqrt{ {( 1 - {1 \over q^2})(q^2 - {1\over {q^2}})} }
\phantom{0}_{s s}{\cal{U}}^{2}_{3}(x) D(y)
\label{38} \\
\phantom{0}_{s s}{\cal{U}}^{3}_{3}(x) B(y) &=&
q^{2}{\phantom{0}_{\sigma s}a(y/x) \over {\phantom{0}_{\sigma s}c(y/x)
} } B(y)
\phantom{0}_{s s}{\cal{U}}^{3}_{3}(x)
-q^{2}{\phantom{0}_{\sigma s}d_-(y/x) \over
\phantom{0}_{\sigma s}c(y/x) }
\phantom{0}_{s s}{\cal{U}}^{2}_{3}(x) D(y)
\nonumber
\end{eqnarray}
We also observe that
\begin{equation}
\label{39}
\phantom{0}_{s s}{\cal{U}}^{1}_{1}(x)\Phi =  \Phi, \quad
\phantom{0}_{s s}{\cal{U}}^{2}_{2}(x)\Phi = q^{-2 L} \phantom{0}_{s s}a(1/x)^L \Phi, \quad
\phantom{0}_{s s}{\cal{U}}^{3}_{3}(x)\Phi = q^{-4 L} \phantom{0}_{s s}b(1/x)^L \Phi
\end{equation}
Then applying the transfer matrix $\phantom{0}_{ss}{\cal T}$ on
the vector $\Psi$ (\ref{29}) and using
eqs. (\ref{37}) and (\ref{39}) we get the eigenvalues of
$\phantom{0}_{ss}{\cal T}$
\begin{eqnarray}
\phantom{0}_{ss}\Lambda(x) &=& q^{2 - 2M} \prod_{i=1}^{M} { \phantom{0}_{\sigma
s}a(x/x_i)\over
\phantom{0}_{\sigma s}c(x/x_i) } \, \, + \, \, q^{-2 L}
{\phantom{0}_{ss}a(1/x)}^L \prod_{i=1}^M
{\phantom{0}_{\sigma s}a(x/x_i)\phantom{0}_{\sigma s}a(x_i/x)
\over {\phantom{0}_{\sigma s}b(x/x_i)
\phantom{0}_{\sigma s}b(x_i/x)} }  \nonumber \\
&+& q^{2(M - 2L -1) } {\phantom{0}_{ss}b(1/x)}^L \prod_{i=1}^M
{\phantom{0}_{\sigma s}a(x_i/x) \over \phantom{0}_{\sigma s}c(x_i/x) }
\label{40}
\end{eqnarray}
We have obtained (\ref{40}) by taking into account only the
first terms on the r.h.s. of eqs. (\ref{37}). All other
terms generate ``unwanted'' contributions and the condition
of their equality to zero yields the BAE (\ref{33}). A
simpler way to recover the BAE from (\ref{40}) is by demanding
that the eigenvalue $ \phantom{0}_{ss}\Lambda(x)  $ (\ref{40})
has no poles at $x = q^{\pm 1/2} x_i $, since $ \phantom{0}_{ss}{\cal T} $
is an analytic function in $x$. Finally, we obtain
the eigenvalues of the Hamiltonian (\ref{22}) from
(\ref{ham}) and (\ref{40})
\begin{equation}
\label{41}
E = \sum_{i=1}^M { {-2 (q^2 - q^{-2})} \over {(x_i^{-1} q^{-1/2} -
x_i q^{1/2}) ( x_i^{-1}q^{3/2} - x_i q^{-3/2} )} }
\end{equation}
In the rational limit $q \rightarrow 1$, this expression
reduces to that obtained by Babujian \cite{bab1} for the
usual periodic case (with appropriate rescaling).

\section*{IV. Highest weight property}

In this section we show that the Bethe vectors are highest
weight vectors with respect to $U_q(sl(2)) $.  
We begin by writing 
\begin{eqnarray}
{\phantom{0}_{\sigma s}R^{\s +}} &=&
\pmatrix{q^{{1 \over 2}h}\, \, &
         0\, \, \cr
         q^{-1 \over 2}(q - q^{-1}) e \, \, &
         q^{{-1 \over 2}h}\, \cr}  \, \, \, ,
\nonumber \\
{\phantom{0}_{\sigma s}R_{\s -}} &=&
\pmatrix{q^{{-1 \over 2}h}\, \, &
         - q^{1 \over 2}(q - q^{-1}) f \, \,\, \, \cr
         0 \, \, &
         q^{{1 \over 2}h}\, \cr}  \, \, \, ,
\label{42}
\end{eqnarray}
where $h$, $e$, $f$ are the $sl(2)$ generators in the spin 1 
representation. Next, defining the constant auxiliary monodromy matrix
as 
\begin{equation}
{\phantom{0}}_{\sigma s}{\cal{U}^-}^{\beta}_{\alpha}
= \lim_{x \rightarrow 0} {\phantom{0}}_{\sigma s}{\cal{U}}^{\beta}_{\alpha}(x) =
( {\phantom{0}}_{\sigma s}{R_{\s -}})^{\beta \, j }_{
\alpha^\prime \, j^\prime} \,
( {\phantom{0}}_{\sigma s}{R^{\s +}})^{\alpha^\prime \, j^\prime }_{
\alpha \, i} \,
\label{43}
\end{equation}
we have from (\ref{42})
\begin{equation}
\label{44}
{C^{\s -}} = q^{-1/2} (q - q^{-1}) q^{{-1 \over 2}h} e.
\end{equation}
The Bethe vectors (\ref{29}) are highest weight vectors if
\begin{equation}
\label{45}
{C^{\s -}} \Psi  = 0
\end{equation}
This can be proven by observing that from the Yang-Baxter algebra (\ref{15})
we can obtain the following relation
\begin{equation}
\label{46}
{C^{\s -}} B(x) = B(x) {C^{\s -}} + (1 - q^{-2}) \biggl( A(x) {D^{\s -}}
- {D^{\s -}} D(x) \biggr)
\end{equation}
which, using the fact that $C^-\Phi=0$, allows us to write
\begin{equation}
\label{47}
{C^{\s -}} \Psi = \sum_{i=1}^{M} Y_i W_i \Phi
\end{equation}
where
\begin{equation}
\label{48}
W_i = B(x_1) B(x_2) \dots B(x_{i-1}) B(x_{i+1}) \dots B(x_M)
.\end{equation}
The $Y_i$ can be computed using eqs. (\ref{30}) and  (\ref{31})
which yields 
\begin{eqnarray}
Y_i &=& (1 - q^{-2}) \, q^{3/2 L}\, \phantom{0}_{\sigma s} a(1/x_i)^{L}
\prod_{j \neq i}^{M} q^{-1}{ \phantom{0}_{\sigma \sigma}a(x_i/x_j)
\over \phantom{0}_{\sigma \sigma}b(x_i/x_j) }
\nonumber \\
&-& (1 - q^{-2}) q^{-L/2} \phantom{0}_{\sigma s}c(1/x_i)^L
\prod_{j \neq i}^M q {\phantom{0}_{\sigma \sigma}a(x_j/x_i) \over
\phantom{0}_{\sigma \sigma}b(x_j/x_i) }
\end{eqnarray}
Because of the BAE (\ref{33}), each of the co-efficients $Y_i$ vanishes
which implies 
$$C^{\s -} \Psi = 0.$$ 
It immediately follows that each of the Bethe states are highest weight
states. By using the $U_q(sl(2))$ lowering operator $f$ we obtain
additional states which will also be eigenstates of the transfer matrix
because of the quantum symmetry of the model.

For generic values of the deformation parameter $q$ it is well
known that the dimensions and weight spectrum of the finite dimensional
irreducible representations of $U_q(sl(2))$ are in 1-1 correspondence
with those of $sl(2)$. Since it is known \cite{2,kir} in the $q=1$ case
that the Bethe states combined with the $sl(2)$ symmetry give a complete
set of states for the model, it should be possible to prove using methods
developed in \cite{KL} that
this is also true for the model described above.

\section*{V. Conclusions}

We have solved a quantum algebra invariant integrable closed spin-1
chain by an algebraic Bethe ansatz approach.
Particularly eigenstates of the model were constructed and their energy
eigenvalues evaluated. A proof of the highest weight property of the Bethe
vectors with respect to $U_q(sl(2))$ was also presented.
A natural extension of this work would be to generalize the results
of the spin-1 chain to corresponding chains of arbitrary spin s.

\section*{Acknowledgements}

J L is supported by an Australian Research Council Postdoctoral
Fellowship.
A F would like to thank H M Babujian for useful discussions
and the Institute
f\"ur Theoretische Physik - FUB for its kind hospitality.
She also thanks DAAD - Deutscher Akademischer Austauschdienst
and FAPERGS -  Funda\c{c}\~{a}o de Amparo \`{a} Pesquisa do Estado
do Rio Grande do Sul for financial support.

\newpage

\end{document}